\documentclass[article]{elsarticle}
\usepackage{lineno,hyperref,amsmath}
\modulolinenumbers[5]
\journal{Energy}
\usepackage{graphicx}
\usepackage[english]{babel}
\usepackage{color,soul}
\usepackage{url}
\definecolor{fgcolor}{rgb}{0.345, 0.345, 0.345}
\usepackage{framed}
\makeatletter
 {\par\unskip\endMakeFramed%
 \at@end@of@kframe}
\makeatother
\definecolor{shadecolor}{rgb}{.97, .97, .97}
\definecolor{messagecolor}{rgb}{0, 0, 0}
\definecolor{warningcolor}{rgb}{1, 0, 1}
\definecolor{errorcolor}{rgb}{1, 0, 0}
\definecolor{red}{rgb}{1, 0, 0}

\usepackage{alltt}
\IfFileExists{upquote.sty}{\usepackage{upquote}}{}
\usepackage[tight,hang,nooneline]{subfigure} 
\subfiguretopcaptrue

\begin{document}

\begin{frontmatter}

\title{Normal behaviour models for wind turbine vibrations: An alternative approach}

\author[os]{Pedro G.~Lind\corref{cor1}}
\ead{pelind@uos.de}
\author[fw]{Luis Vera-Tudela}
\author[fw]{Matthias W\"achter}
\author[fw]{Martin K\"uhn}
\author[fw]{Joachim Peinke}

\cortext[cor1]{Corresponding author}
\address[os]{Institut f\"ur Physik, Universit\"at Osnabr\"uck,
         Barbarastrasse 7, 49076 Osnabr\"uck, Germany}
\address[fw]{ForWind - Center for Wind Energy Research, Institute of Physics, Carl von Ossietzky University of Oldenburg, K\"upkersweg 70, 26129 Oldenburg, Germany}


\begin{abstract}
To monitor wind turbine vibrations, normal behaviour models are built to predict tower top accelerations and drive-train vibrations. Signal deviations from model prediction are labelled as anomalies and are further investigated.
In this paper we assess a stochastic approach to reconstruct the $1$ Hz tower top acceleration signal, which was measured in a wind turbine located at the wind farm Alpha Ventus in the German North Sea. We compare the resulting data reconstruction with that of a model based on a neural network, which has been previously reported as a data-mining algorithm suitable for reconstructing this signal.
Our results present evidence that the stochastic approach outperforms the neural network in the high frequency domain ($1$ Hz). Although neural network retrieve accurate step-forward predictions, with low mean square errors, the stochastic approach predictions better preserve the statistics and the frequency components of the original signal, remaining high accuracy levels.  
The implementation of our stochastic approach is available as open source code and can easily be adapted for other situations involving stochastic data reconstruction.
Based on our findings we argue that such an approach could be implemented in signal reconstruction for monitoring purposes or for abnormal behaviour detection. 
\end{abstract}


\end{frontmatter}

\section{Introduction}

As the number of wind turbines installed worldwide increases, the reduction of operational and maintenance cost by enhanced surveillance gains on importance \cite{Yang2014}. Recent reviews on condition monitoring of wind turbines \cite{Kusiak2013, Wang2014a, Tchakoua2014}, recognize the importance that a reliability based maintenance system has on reducing the cost of energy. Although it is possible to monitor wind turbine functioning with 10-minute average SCADA data, part of the information is lost when data is averaged; a mixed system, combining SCADA data and conventional high frequency sensors is expected in the
future \cite{Wang2014a}.

Zaher et al \cite{Zaher2009} described the lack of failure records as the main challenge for early failure detection, which emphasizes the importance of normal behaviour modelling. Such models are built empirically, i.e. using only measured signals, and refer to the expected value the signal should have based on its own historical records. An anomaly in the signal is detected when the error on its prediction, made by the normal behaviour model, increases over several subsequent values. 

Previous investigations have built normal behaviour models with data-mining algorithms on SCADA data. Most of them are based on $10$-minute data \cite{newref}. Cruz Garcia and co-workers \cite{CruzGarcia2006} presented their application to detect anomalies in wind turbine gearboxes. In Ref.~\cite{Zaher2009} the authors proposed a framework for their combination into a single user interface to facilitate their use, while Ref.~\cite{Schlechtingen2011} shows that a normal behaviour model based on neural networks (NN) outperforms a regression based one. But some scientists also evaluate data with higher sampling frequency, Kusiak and co-workers \cite{Kusiak2012a} analysed the performance of various NN algorithms to detect bearing faults on wind turbines. In Refs.~\cite{Kusiak2010h} and \cite{Zhang2012b}, they evaluated wind turbine vibrations in time domain with $10$-second sampling signals from drive-train vibrations and tower top accelerations.  

In contrast to existing methods, signals can also be reconstructed based on a stochastic approach, which has not been assessed so far for the creation of normal behaviour models. In a previous work \cite{Lind2014}, we demonstrated its value to estimate fatigue loads for wind turbines based on reconstructed signals.
Moreover, this stochastic approach was already combined with NN models \cite{Russo2013} for improving forecast of air quality.
The objective in this paper is to assess the stochastic
approach as basis for normal behaviour models. We evaluate the performance of such model to monitor vibrations on wind turbines and compare it versus one based on a NN, which was found the most suitable NN model to reconstruct the tower top acceleration \cite{Zhang2012b}. We focus our assessment on the similarities and differences between both approaches, we then discuss our findings with respect to the limitations imposed by the specific case evaluated.

We start in Sec.~\ref{sec:data} by describing in detail the data used in this investigation, obtained from one 
wind turbine in an offshore wind farm.
In Sec.~\ref{sec:methods} both the NN model and our stochastic approach are introduced; then, the results obtained from them are properly plotted in the figures and tables below. In Sec.~\ref{sec:results} we compare the results of both approaches when reconstructing one month of the tower acceleration measured at one turbine. Section \ref{sec:conclusions} concludes the paper.

\section{Data description}
\label{sec:data}

Wind turbine vibrations can easily be assessed monitoring two standard operational signals: drive-train and tower top acceleration \cite{Zhang2012b}. In this paper, we limited the analysis to the second one to assess the reconstruction of signals with both the neural network model and the stochastic approach.  

The data analysed in this paper correspond to measurements carried out in one 5MW Senvion wind turbine AV-04 of the offshore wind farm Alpha Ventus \cite{Mueller2016}. Alpha Ventus is the first offshore wind farm in Germany, located at Borkum West in the North Sea, $54.3^o$N-$6.5^o$W. 

The turbine is sujected most of the time to undisturbed inflow due to its position in the first row of turbines, located at the western past of the wind farm.
The support structure of the turbine combines a rather stiff and hydrodynamically transparent jacket structure with a tubular tower on top.
In such a design the tower top acceleration signal is dominated by the dynamic response on aerodynamic excitation mainly at the first bending natural frequency of the system and at multiples of the blade passing frequency. Wave excitation plays only a minor role.

Data include all wind speed and tower top acceleration values recorded in October and November 2014, regardless of the operational and wind flow conditions. Therefore, $29$ days were available in October ($2 555 805$ values) and $21$ days in November ($1 859 179$ values). We only removed incorrect measurements from the data collected, either because they were identified by the recording system with a flag ('99999' in our case) or because their values were out of reasonable physical values, namely $x>\bar{x}+5\sigma$. Wind speed and tower acceleration values were available with sampling ratios of $1$ Hz and $50$ Hz respectively, thus the last one was down-sampled to $1$ Hz in order to complete the analysis.

\begin{figure}[t] 
\includegraphics[scale=0.36]{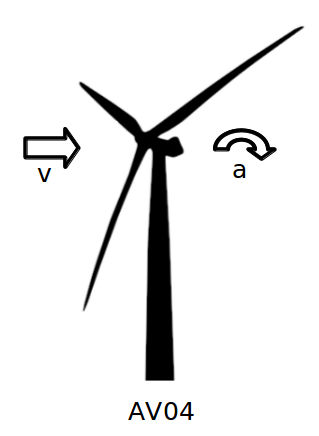}%
\includegraphics[scale=0.32]{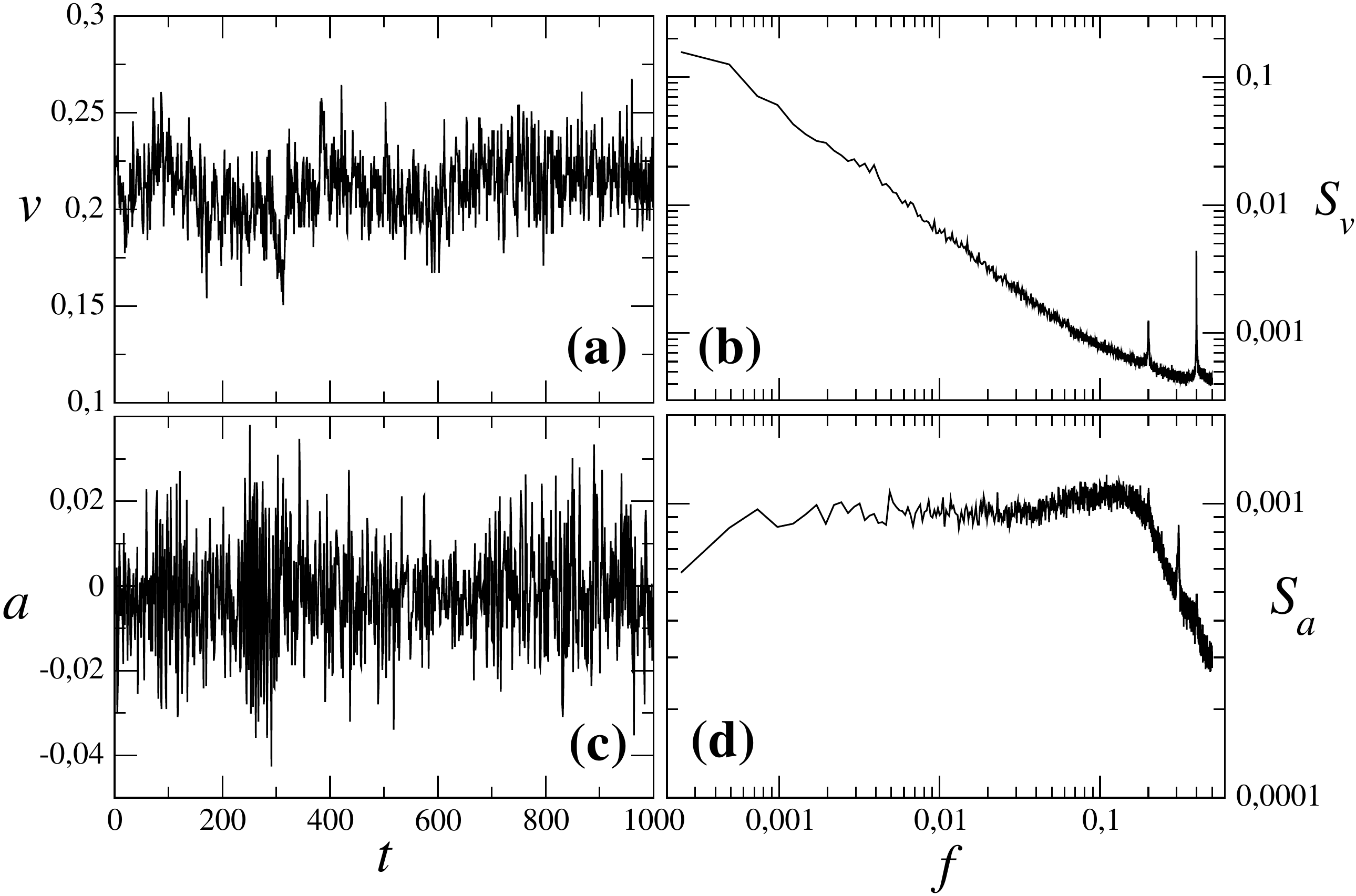}
\centering
\caption{Examples of 
         {\bf (a-b)} wind speed and 
         {\bf (c-d)} tower acceleration signals in Alpha Ventus wind farm. 
         In the left one ilustrates the time-domain of both 
         quantities while in the right the frequency domain 
         is shown for the full datasets of October 2014.
         All data was normalized to fulfill all confidentiality
         protocols (see text). Frequency units are Hertz and time is
         given in seconds. Both wind velocity and tower acceleration
         are normalized to maximal observed values.}
\label{fig01}
\end{figure}

Wind speeds were measured by a cup anemometer included in the meteorological station located on top of the nacelle. Tower top longitudinal accelerations were recorded by an accelerometer positioned at the bottom of the nacelle, near its connection to the upper part of the tower. Figure \ref{fig01} illustrates the characteristics of the data in time and frequency domain. 
\begin{figure}[t] 
\includegraphics[scale=0.5]{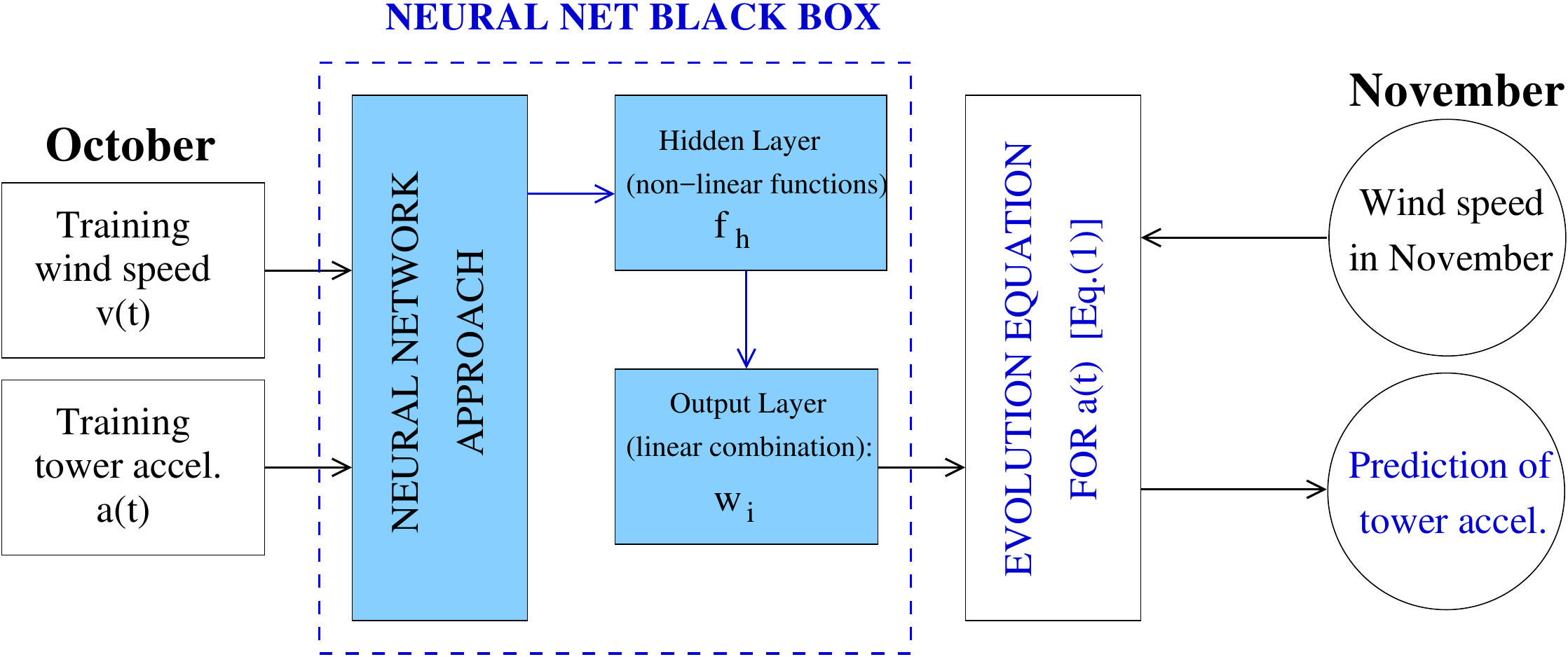}
\centering
\caption{Schematic illustration of the neural network (NN)
 approach, indicating the development and test of neural networks for
 signal reconstruction.  Here the particular case of tower
 acceleration reconstruction 
 is given.}
\label{fig02}
\end{figure}

To protect sensitive manufacturer information we normalized the
values. Wind speeds recorded included the whole operational range of
the wind turbine. Furthermore, we divided measurements in two sets
according to different purposes: data collected in October were used
to estimate the parameters in each model, while data collected in
November were used to test each model performance. 

\section{Methods and results}
\label{sec:methods}

Each approach to construct a normal behaviour model is introduced in a concise manner. Where convenient, we provide references to literature with extensive details about neural networks and the stochastic approach. At the end of the section, we describe the performance metrics utilised for the assessment.

Zhang and Kusiak \cite{Zhang2012b} evaluated a pool of potential input variables to predict tower top acceleration signal with a wrapper algorithm \cite{Borovicka2012}, which includes the predictor model to search for the variables that reduce prediction error ({\it a posteriori} approach). They found wind speed, tower acceleration and wind direction relevant. Since our evaluation focused on the approaches themselves, we limited the pool of potential input variables to a single external signal (wind speed) and previous values of the tower top acceleration to have the same number of inputs for each approach.

\subsection{Neural networks: a deterministic approach}

Neural networks transform input variables with a linear coefficient and a non-linear function. This is repeated, in a chain, over a variable number of layers, where each output layer becomes the input of the next one. The output of the last layer is the final prediction made for the target value. In this analysis, we utilised a non-linear auto-regressive with exogenous inputs (NARX) neural network, which is a recurrent network, i.e.~it includes one or more feedback loops. Thus, the approach followed to estimate the predicted value of the tower top acceleration, 
given by 
\begin{equation} 
\hat{a}(t) = f[a(t-\Delta t),\dots,a(t-n_a \Delta t),v(t-\Delta t),\dots,v(t-n_v \Delta t)]+e(t)  ,
\label{eq:generic}
\end{equation}
involves the use of its $n_a$ previous values ($a(t- \Delta t), \cdots, a(t-n_a \Delta t)$) and those of the $n_v$ previous wind speeds ($v(t- \Delta t), \cdots, v(t-n_v \Delta t)$) through a smooth unknown function $f()$. An extra random component $e(t)$ represents a random error, which has a zero mean and is independent of $v$ and $a$ \cite{Hastie2013}. The time-dependent predicting function $\hat{a}(t)$ can be taken as an evolution equation. 

To create the baseline normal behaviour model we used a NN to approximate $f()$, which is a recursive non-linear transformation of its inputs. Zhang and Kusiak \cite{Zhang2012b} demonstrated that NNs were the most suitable algorithm to create a normal behaviour model for the tower top acceleration of a wind turbine when compared with neural network ensemble, boosting regression trees, support vector machine, random forest with regression, standard classification and regression tree and K-nearest neighbour. In our investigation, we used a NN with two layers (two transformations). The first one (hidden layer) consisted of $n_h$ neurons, which apply non-linear transformations via a sigmoid function; the second one (output layer) was formed by a single neuron, which performed an extra linear transformation, with given weights $w_i$. 

To optimize the neural network we can also include a finite number of feedback loops for previous values of the tower top accelerations ($a, n_a$) and wind speeds ($v, n_v$), thus we have three parameters to determine ($n_h, n_v, n_a$).  

First, we divided data from October 2014 using the hold out method \cite{Borovicka2012} in sub-sets for model training (70\%),
validation (20\%) and test (10\%). Then, we selected the number of neurons in the hidden layer ($n_h$), the number $n_v$ of input delays for the wind speed $v$ and the number $n_a$ of output delays for the tower top acceleration $a$ with a wrapper algorithm \cite{May2011}, which searched for optimum prediction in the ranges $n_h,n_a,n_v \in \{ 1, \cdots, 50 \}$.  

The optimization of the unknown function $f$ was set to minimize mean squared error 
\begin{equation} \label{eq:MSE}
\mathit{MSE} = \frac{1}{n} \sum_{i=1}^n (\hat{y}_i - y_i)^2 ,
\end{equation}
using the Levenberg-Marquardt algorithm. As a result, the architecture of the NARX network consisted on $25$ neurons in the hidden layer ($n_h = 25$), three input delays ($n_v = 3$) and one output delay ($n_a = 1$). 

Figure \ref{fig02} depicts the development and utilization of the NN model investigated. From left to right, first we used wind speeds and tower top accelerations collected in October 2014 to define the constitutive components of the model, which is named a neural net black box in Fig.~\ref{fig02}. Once the evolution equation $\hat{a}(t)$ was defined, we used wind speeds collected in November 2014 to predict the tower top acceleration signal, as well as to evaluate the performance of the model. 
\begin{figure}[t] 
\includegraphics[scale=0.5]{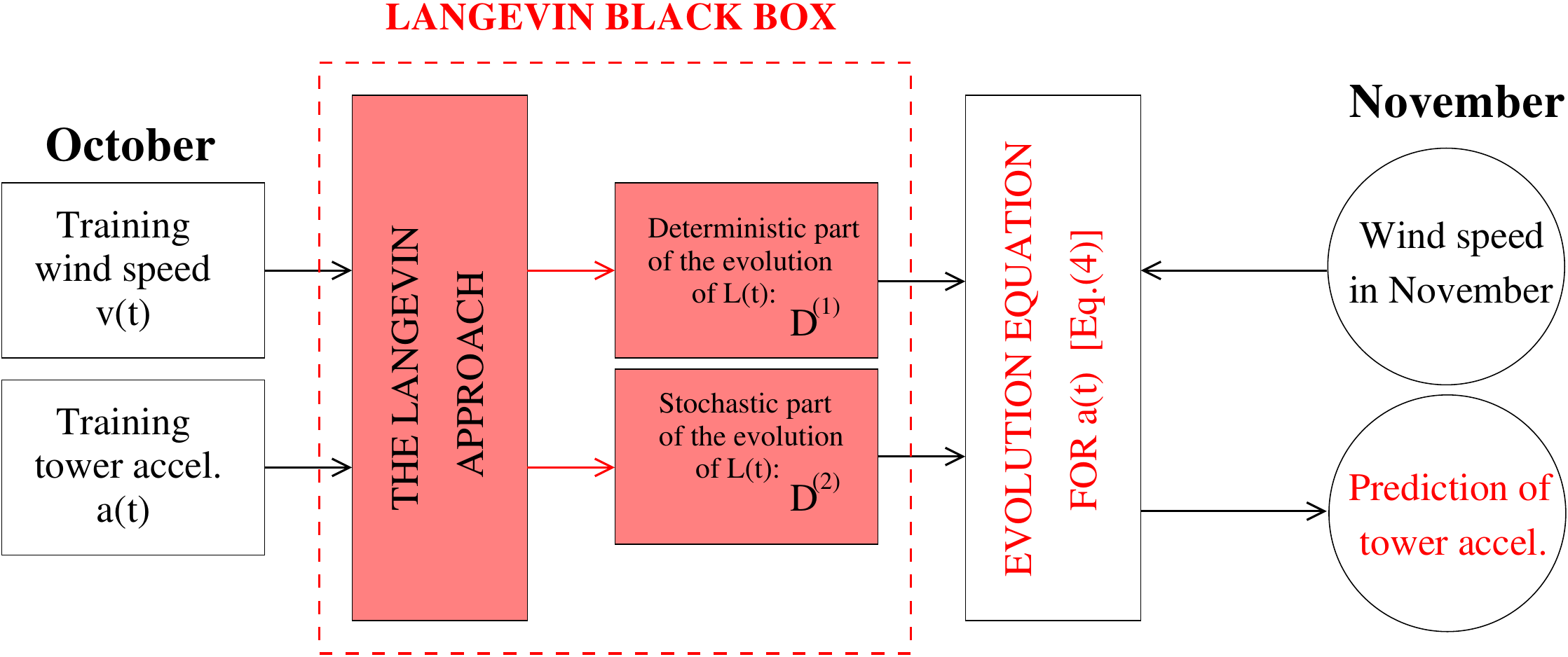}
\centering
\caption{Schematic illustration of the Langevin (stochastic) 
approach.
Here the particular case of tower acceleration reconstruction at Alpha
Ventus is given.}
\label{fig03}
\end{figure}

\subsection{Stochastic approach: the Langevin model}

The Langevin model is a framework developed from the pioneering 
work by Peinke and Friedrich in 1997 \cite{friedrich97,siegert1998}, which 
consists of a direct method for extracting the evolution 
equation of stochastic series of measurements.
Several applications were proposed and
developed, e.g.~in turbulence modelling, in medical EEG monitoring
and in stock markets. 
See Ref.~\cite{physrepreview} for a review.
In the context of wind energy, this framework has shown the ability
for predicting power curves of single wind 
turbines as well as
of equivalent power curves for entire wind farms, and
also to properly reproduce the increment statistics of
power and torque in single wind
turbines from wind speed measurements \cite{muecke}.

\begin{figure}[htbp] 
\includegraphics[scale=0.5]{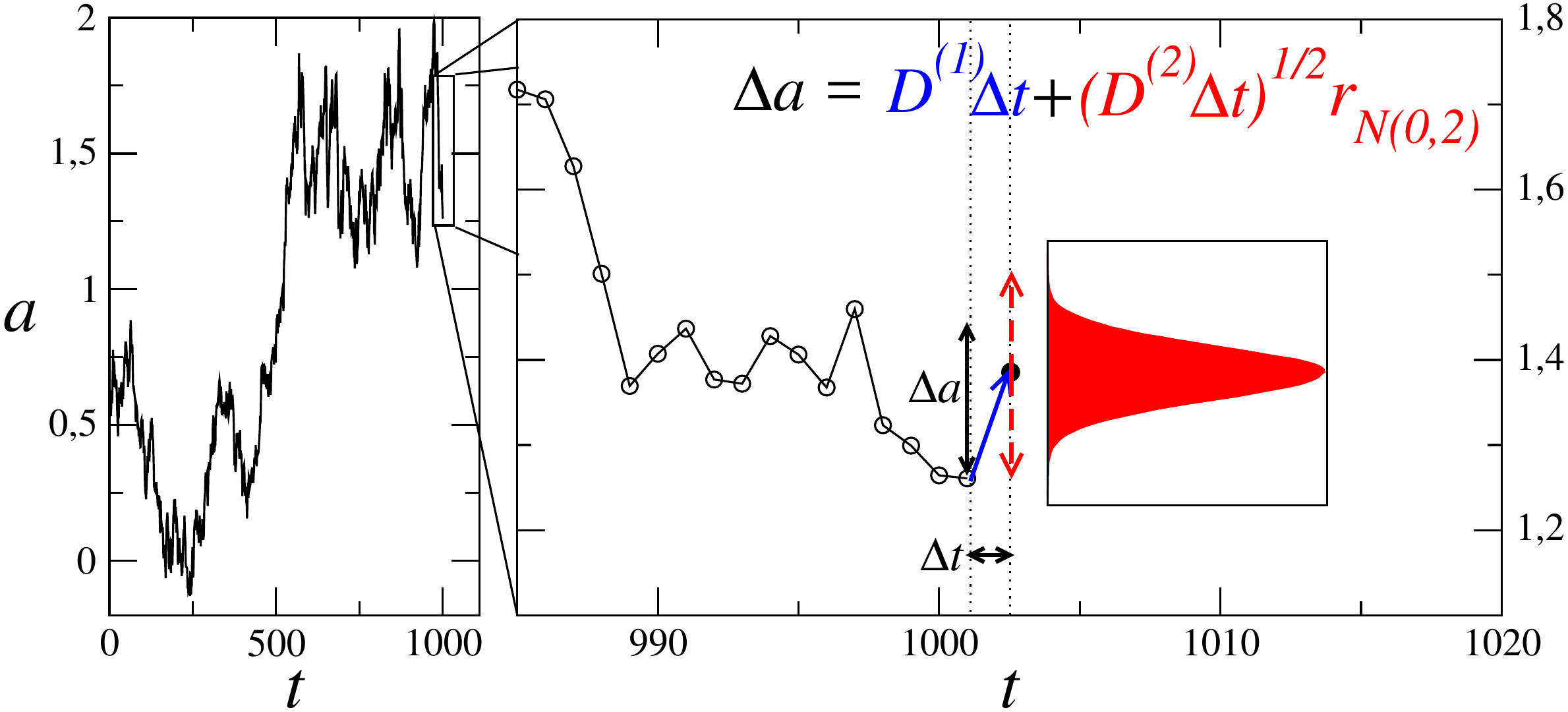}
\centering
\caption{Illustration of the stochastic approach, while integrating 
  the evolution equation (\ref{LangVectDisc}),
  composed by its two
         contributions, the deterministic contribution $D^{(1)}$
         which governes the tendency of the evolving observable
         (blue), and the stochastic fluctuations accounted by
         $D^{(2)}$ (red) added to the deterministic part.}
\label{fig04}
\end{figure}

Figure \ref{fig03} depicts the development and utilization of the
stochastic approach investigated. Our stochastic framework, instead of
computing non-linear functions and weights, which optimize a set of
output compared to input data, retrieves two single functions of the
variables involved. One of such functions, below symbolized by
$D^{(1)}$, governs the deterministic contribution for the time
variation of the output variable, while the other function, $D^{(2)}$,
accounts for the stochastic fluctuations that include all the
non-observable degrees of freedom present in the system. See the
illustration in Fig.~\ref{fig04}.  

Having wind speeds $v(t)$ and tower accelerations $a(t)$, the following evolution equation for the later quantity is used:
\begin{equation}
\frac{da}{dt}=D^{(1)}(a,v)+\sqrt{D^{(2)}(a,v)}\Gamma_t,
\label{LangVect}
\end{equation}
with $\Gamma_t$ representing a Gaussian $\delta$-correlated white noise,
i.e.~$\langle \Gamma_t\rangle = 0$ and 
$\langle \Gamma_t\Gamma_{t^{\prime}}\rangle = 2\delta(t-t^{\prime})$.

The reconstruction of the tower
acceleration $a(t)$ follows therefore 
directly from the stochastic integration of Eq.~(\ref{LangVect})
which yields: 
\begin{equation}
a(t+\Delta t) = a(t)+D^{(1)}(a,v)\Delta t+ \sqrt{D^{(2)}(a,v)\Delta t} \; r_{N(0,2)} ,
\label{LangVectDisc}
\end{equation}
where $r_{N(0,2)}$ is a random number from a Gaussian distribution 
with zero mean and variance equal to two.
Details are illustrated in Fig.~\ref{fig04}
and can be found in Ref.~\cite{physrepreview}.

Function $D^{(1)}$ is also called drift function, while function
$D^{(2)}$ is typically named as the diffusion function.
Since the drift and diffusion functions have a physical interpretation,
one could apply the model in Eq.~(\ref{LangVect}) to a particular system
and define {\it ad hoc} the functional shape of both functions from
physical reasoning. Next we explain how to compute these both functions.

The derivation is performed through the computation of the
corresponding first and second conditional moments \cite{Lind2014},
respectively
\begin{subequations}
\begin{eqnarray}
M^{(1)}_{v^{\ast}}(a,\tau) &=& \left\langle a(t+\tau)-a(t)\right\rangle
|_{a(t)=a,v(t)=v^{\ast}} , \\
M^{(2)}_{v^{\ast}}(a,\tau) &=& \left\langle (a(t+\tau)-a(t))^2 \right\rangle
|_{a(t)=a, v(t)=v^{\ast}} .
\end{eqnarray}
\label{condmoments}
\end{subequations}
where $\langle \cdot \rangle |_{X(t)=x}$ indicates the average
over the full time series, whenever $X(t)$ takes the value $x$,
defined within an interval.

\begin{figure}[t]
\begin{center}
\includegraphics[width=0.38\textwidth]{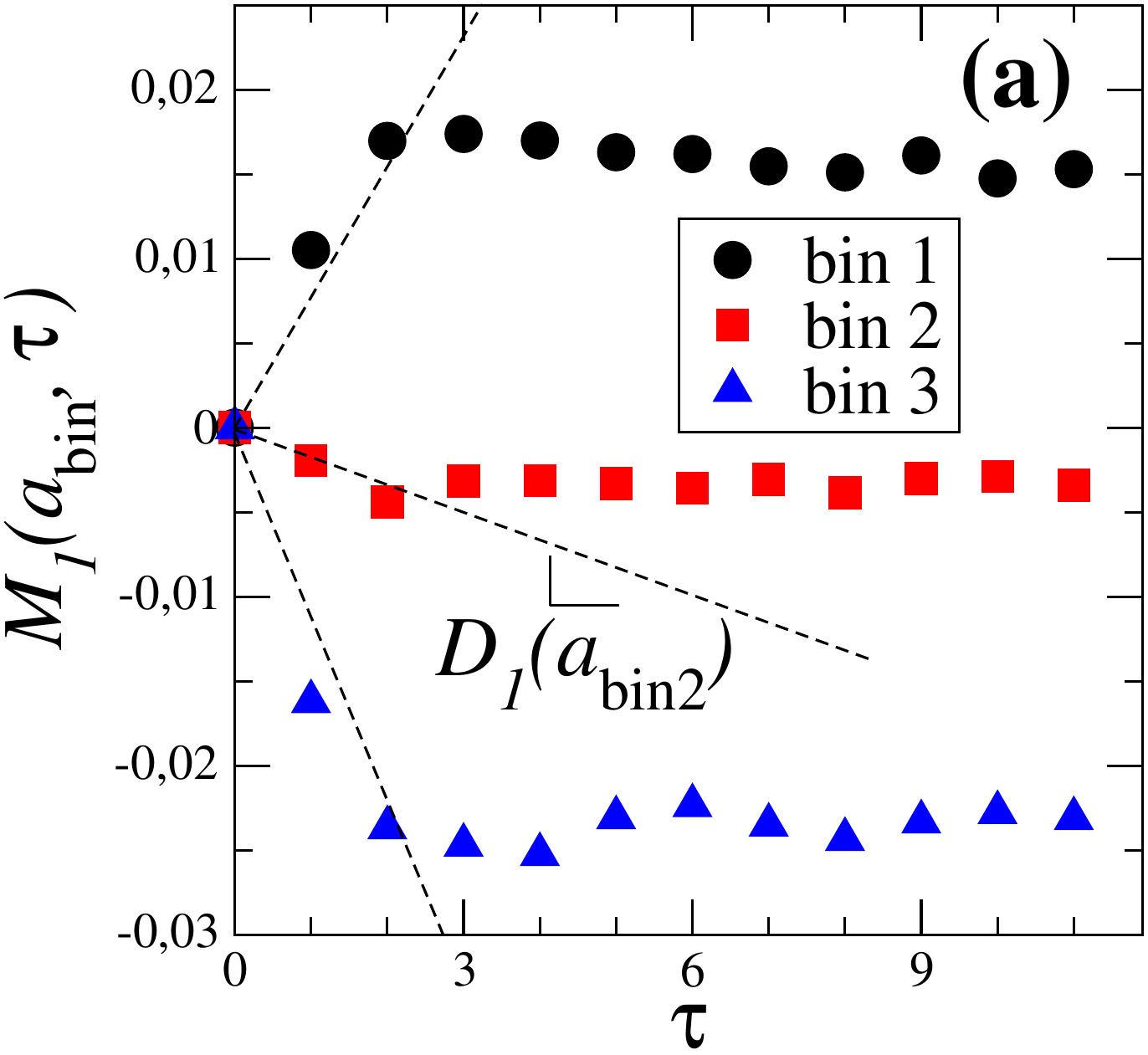}%
\hspace{0.5cm}%
\includegraphics[width=0.55\textwidth]{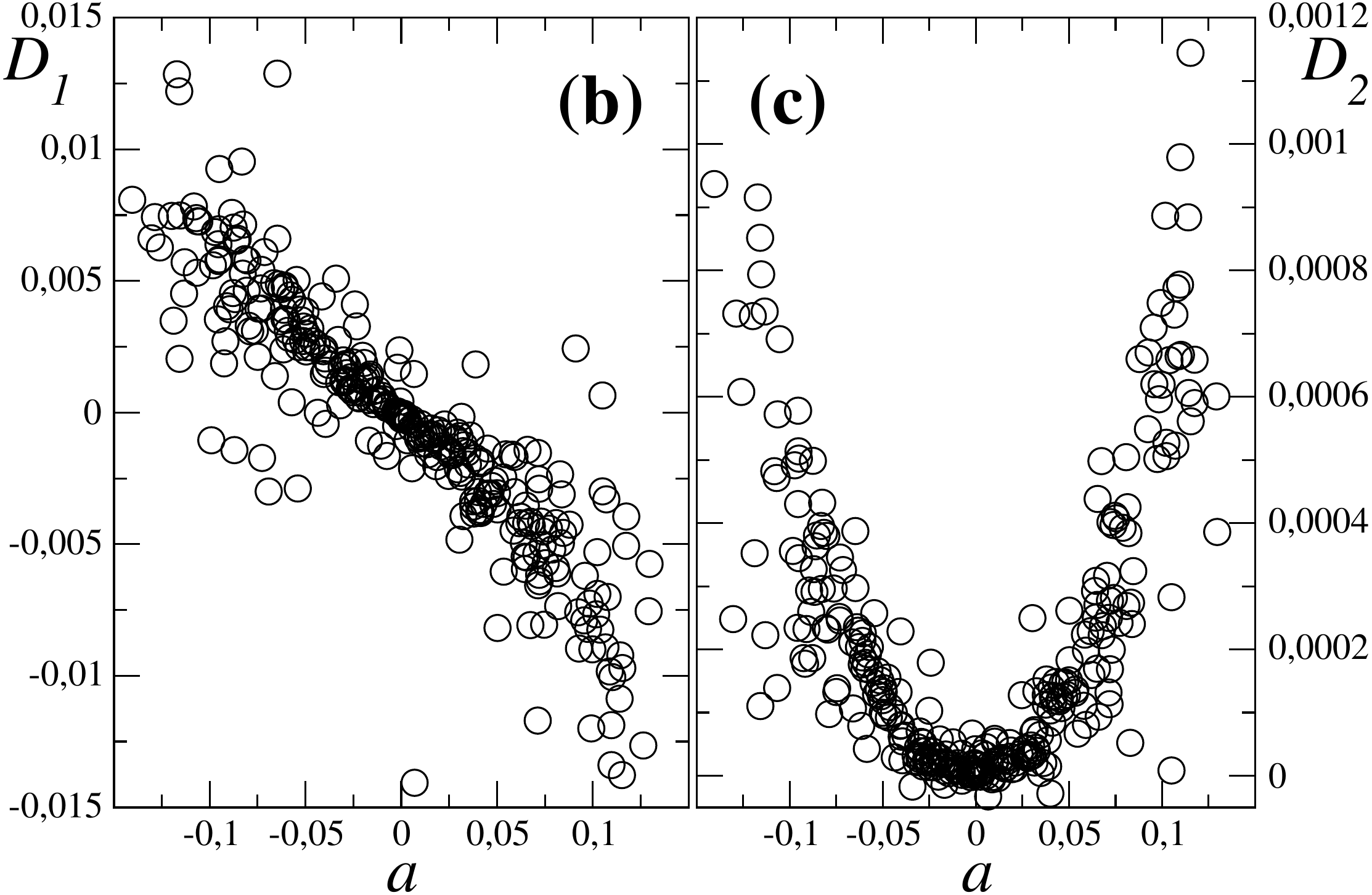} 
\caption{\protect 
         Illustration of 
         {\bf (a)} the first conditional moment for three different
              bin values of the tower acceleration as defined 
              in Eqs.~(\ref{condmoments}). 
              Here $v^{\ast}$ is given by the average velocity found
              during October. 
         Numerical result for 
         {\bf (b)} the drift $D^{(1)}(a,v)$ and 
         {\bf (c)} the diffusion $D^{(2)}(a,v)$ in the Langevin equation 
         given by Eq.~(\ref{DefCoefKM}),
         plotted as function of $a$ alone, i.e.~they are projected at the
         $a$-axis to emphasize the linear and quadratic dependency of
         $D^{(1)}$ and $D^{(2)}$ respectively for the largest range of
         acceleration values.}
\label{fig05}
\end{center}
\end{figure}

Figure \ref{fig05}a shows the first 
conditional moment for different values of the tower acceleration.
Since it can be shown 
that drift ($D^{(1)}$) and diffusion functions ($D^{(2)}$) in 
Eq.~(\ref{LangVect}) are, apart from a
multiplicative constant ($1/k!$), the derivative with respect to the 
time-gap $\tau$ of the first and second conditional moments 
respectively, we can define both functions directly from the
conditional moments, namely:
\begin{equation}
D^{(k)}(a,v)=\lim_{\tau\rightarrow0}\frac{1}{k!}\frac{M_v^{(k)}(a,\tau)}{\tau} .
\label{DefCoefKM}
\end{equation}

In other words, by taking the slope of the linear regression for each 
conditional moment one arrives to the corresponding value of function
$D^{(1)}$ and $D^{(2)}$.
Indeed, as sketched in Fig.~\ref{fig05}a, within a sufficiently low
range of $\tau$ values, here three time-steps of the set of
measurements, the conditional moments $M^{(1)}$ and $M^{(2)}$ 
depend linearly on $\tau$.  
Figures \ref{fig05}b and \ref{fig05}c show 
both drift and diffusion functions, ($D^{(1)}$) and ($D^{(2)}$) respectively, for a range of
$a$-values at different velocities.
Notice that while the diffusion function shows a quadratic dependence
on the tower acceleration, the drift function exhibits a cubic dependence
with a dominant linear term having one single fixed point at $a=0$ which
corresponds to the equilibrium position of tower vibrations.

It is important to stress that our framework is based on the
assumption that the observable follows a Markov process, which means
nothing else than that the noise $\Gamma_t$ is $\delta$-correlated as
mentioned above.
Though, one advantage of this Langevin data reconstruction
is that it also works in some cases where the Markov
test fails \cite{philip}.
The full implementation in one- and two-dimensions of this
approach is already public available
\cite{routine}.
Improvements on the noise term are beyond this paper, but were
already addressed previously\cite{Lehle2011,anvari}.

\subsection{Performance evaluation}

\begin{figure}[t] 
\includegraphics[scale=0.4]{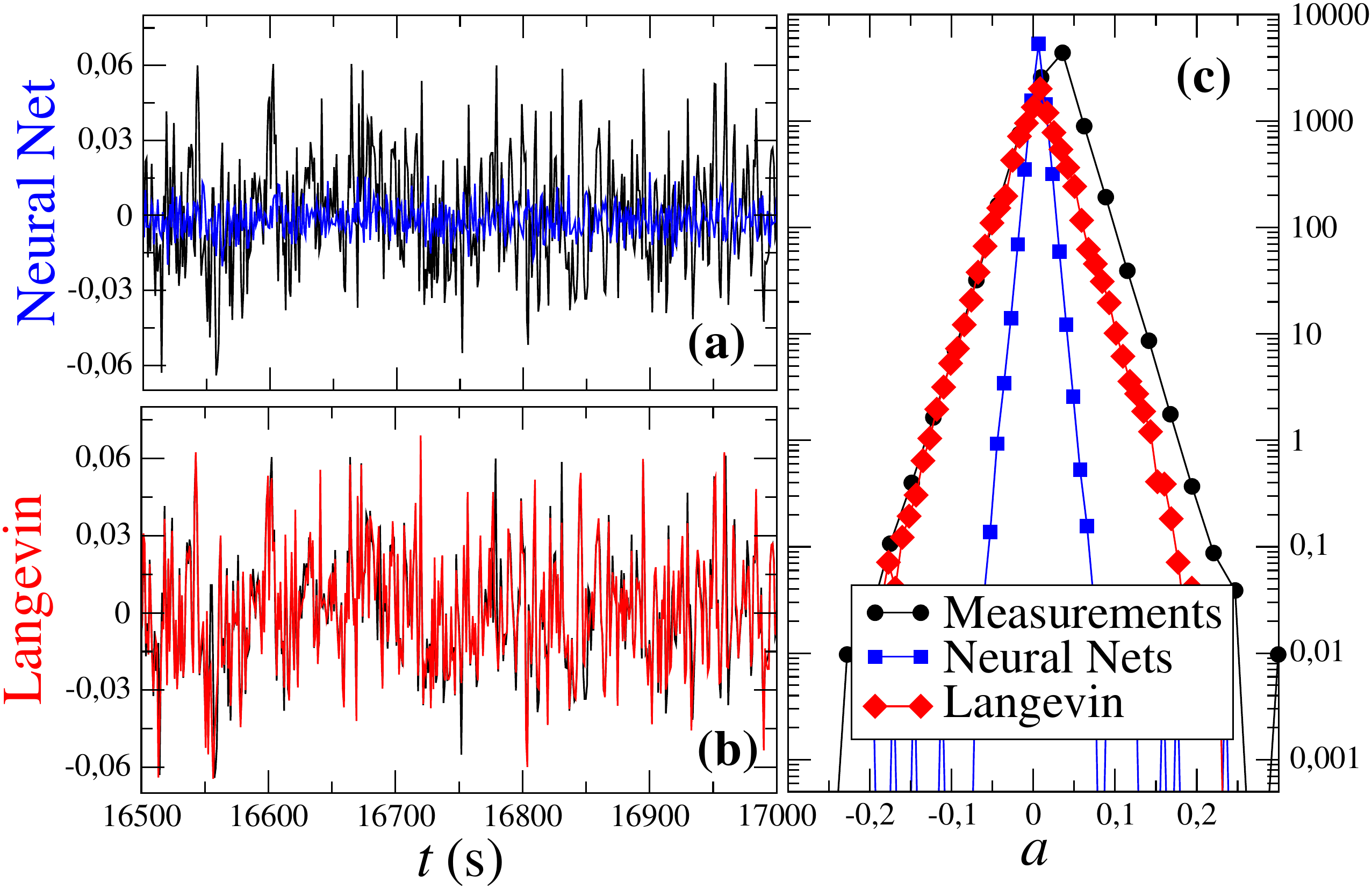}
\centering
\caption{Sample of data series of tower acceleration from
         measurements at AV04 in Alpha Ventus and 
         the acceleration reconstruction models using
         {\bf (a)} NARX neural networks and
         {\bf (b)} the Langevin model.
         In both cases, one also plots (black lines)
         the corresponding series of measurements for better
         comparison.
         The corresponding value distribution of these series is
         displayed in
         {\bf (c)} with symbols for the full month of November 2014.}
\label{fig06}
\end{figure}

To compare both approaches, we consider the distribution of all values
predicted from each model for November 2015, in particular their four
first statistical moments, namely the mean, the standard deviation,
the skewness and the kurtosis. 
Notice that, while mean and standard deviation are sufficient for
characterizing the distribution of Gaussian processes, in general,
higher-order moments should be checked for ascertaining if the process
is non-Gaussian or not. 
The corresponding results are shown in Fig.~\ref{fig06} and
Tab.~\ref{tab:moments}. 

Additionally,
to evaluate the prediction of both models we use four different metrics,
namely the mean absolute error
\begin{equation} \label{eq:MAE}
\mathit{MAE} = \frac{1}{n} \sum_{i=1}^n \vert \hat{y}_i - y_i\vert ,
\end{equation}
the standard deviation of the absolute error
\begin{equation} \label{eq:SDofAE}
\mathit{SDofAE} = \sqrt{ \frac{1}{n} \sum_{i=1}^n \left ( \vert\hat{y}_i - y_i\vert - \frac{1}{n} \sum_{i=1}^n \vert\hat{y}_i - y_i\vert \right ) ^2 }  ,
\end{equation}
the mean square error (already presented in Sec.~\ref{sec:methods},
Eq.~(\ref{eq:MSE})) and the standard deviation of the square error 
\begin{equation} \label{eq:SDofSE}
\mathit{SDofSE} = \sqrt{ \frac{1}{n} \sum_{i=1}^n \left [ (\hat{y}_i -
  y_i)^2 - \sum_{i=1}^n (\hat{y}_i - y_i)^2 \right ] ^2 }  ,
\end{equation}
where $\hat{y}_i$ denotes the $i$th measured value of the property
$y$, and $y_i$ the corresponding modelled value, either with NN or
with the Langevin model.
\begin{table}[t] 
\centering
\begin{tabular}{p{3.5cm} cccc}
\hline
Signal & Mean & Std. Dev. & Skewness & Kurtosis \\
\hline
Measurements   & -2.07E-3 & 24.7E-3 &  11.6E-3 & 5.07 \\ 
NN             & -2.15E-3 & 7.39E-3 & -30.1E-3 & 6.31 \\  
Langevin model & -1.71E-3 & 25.6E-3 & -3.59E-3 & 2.37 \\ 
\hline 
\end{tabular} 
\caption{\protect
         First four statistical moments of the value distributions
         shown in Fig.~\ref{fig06}d for the normalised measurements
         and the reconstructed signals with each one of both models.} 
\label{tab:moments}
\end{table}
\begin{table}[t] 
\centering
\begin{tabular}{p{3.5cm} cccc}
\hline
Signal & MAE & SDofAE & MSE & SDofSE \\
\hline
NN & 0.031 & 0.032 & 2.0E-3 & 65.6E-3 \\  
Langevin model & 0.027 & 0.031 & 1.6E-3 & 0.7E-3 \\ 
\hline 
\end{tabular} 
\caption{\protect
         Performance of both models, using the metrics defined in
         Eqs.~(\ref{eq:MSE}),(\ref{eq:MAE})-(\ref{eq:SDofSE}), namely
         the mean of the absolute error, Eq.~(\ref{eq:MAE}),
         its standard deviation, Eq.~(\ref{eq:SDofAE}),
         the mean square error, Eq.~(\ref{eq:MSE}) and
         its standard deviation, Eq.~(\ref{eq:SDofSE}).}
\label{tab:metrics}
\end{table}

References \cite{Zhang2012b} and \cite{Zhang2012a} reported these
metrics to select the best data-mining technique and to derive models for
the normal behaviour of wind turbine vibration and to detect faults in
wind turbine gearboxes.  
Moreover, they have been commonly used to evaluate the accuracy of
models utilised for regression analysis in wind energy
applications \cite{Kusiak2012a, Kusiak2010h, Zhang2012b, Zhang2012a}.  
The correspondig results comparing the NN and the
stochastic approach are summarized in Tab.~\ref{tab:metrics}.  

\begin{figure}[t]   
\includegraphics[scale=0.4]{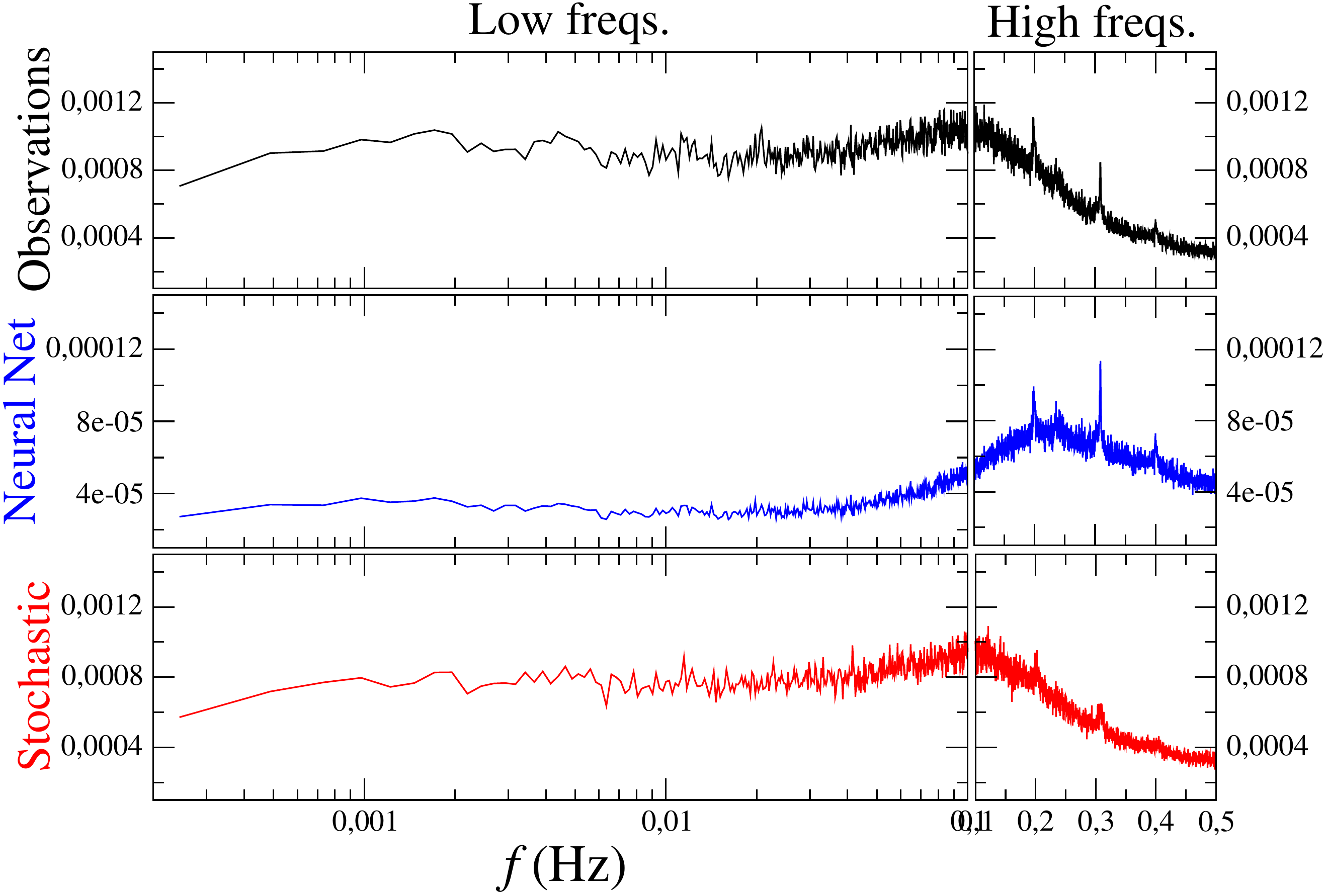}
\centering
\caption{\protect
         Spectrum of original signal (top) and the corresponding
         reconstructed signals, one using the neural
         network (middle) and the other using our proposed stochastic
         approach (bottom). Notice that the high frequency domain is
         plotted in a linear frequency scale, while the low frequency
         domain is plotted in the logarithmic scale.} 
\label{fig07}
\end{figure}

\begin{figure}[htbp] 
\includegraphics[scale=0.4]{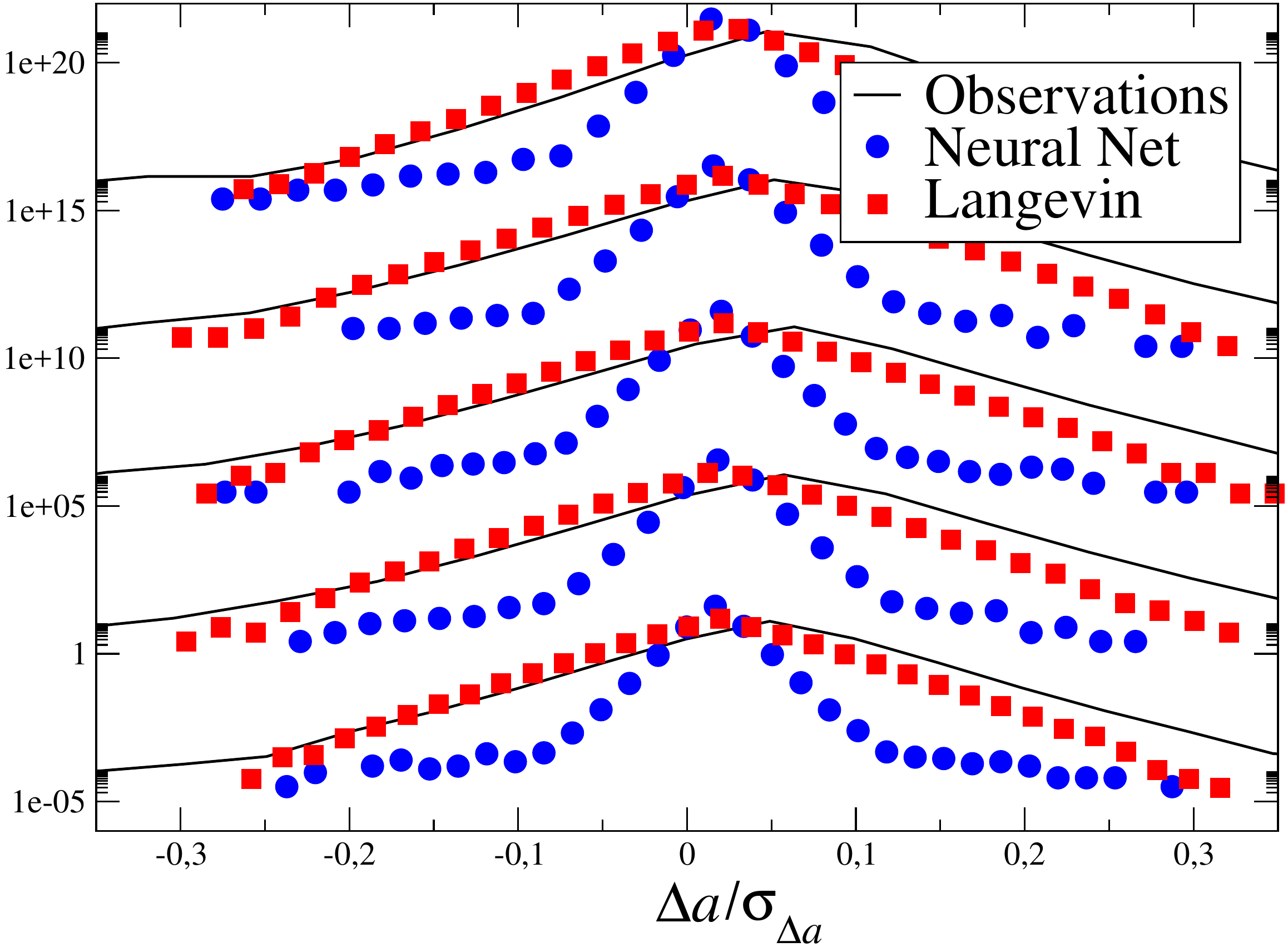}
\centering
\caption{\protect
  Two-point statistics of the tower acceleration (lines)
  and the corresponding reconstructed
         signals, i.e.~value distributions of $\Delta
         a(t)=a(t+\tau)-a(t)$. From top to bottom one has
         $\tau=1,2,4,8$ and $16$ seconds. The vertical shift of the
         distribution is for better visualization.} 
\label{fig08}
\end{figure}

Finally, we also analyse the temporal correlations of the data.
Therefore we use two methods, namely we calculate the power spectra
and analyse the statistics of the temporal increments
\begin{equation}
\Delta a (t,\tau) = a(t+\tau)-a(t) ,
\end{equation}
where $\tau=n\Delta t$ is an integer number $n$ of consecutive
time-steps $\Delta t$.
Note that the second moment of the increments $\Delta a$, given by
$\langle (\Delta a)^2 \rangle$, corresponds to the power
spectrum. 
The power spectrum of the reconstructed signals are plotted in
Fig.~\ref{fig07}, showing how the frequency components are
reconstructed. 
The statistics of the increments is given in Fig.~\ref{fig08}, from
which we will ascertain how good the model retrieves the evolution of
the process throughout the succession of
measurements \cite{physrepreview}. 

\section{Discussion: comparing both approaches}
\label{sec:results}

To evaluate the suitability of the stochastic approach, described
above, for modelling normal behaviour of wind turbine vibrations, 
we compare it with the neural network \cite{Zhang2012b}. 

Figures \ref{fig06}a and \ref{fig06}b
show normalized samples of the original
tower top acceleration and its two reconstructions. Already in these
plots it is clear the stronger resemblance between the reconstructed
data from the stochastic approach and the set of tower acceleration
measurements.
While the NN reconstruction keeps the average value of the tower
oscillations, their range of amplitudes is not as well reproduced
as with the stochastic approach.

This feature is more evident in the plot of 
Fig.~\ref{fig06}c: both the stochastic and the NN models
retrieve the approximate distribution of values of the original
signal (symbols), but
the full range of the tower acceleration values observed in
measured distribution (bullets) is reproduced by the distribution
obtained from the Langevin model (diamonds) but not from the NN
model (squares).

Furthermore, the empirical distribution is clearly non-Gaussian,
showing exponential tails.
Table \ref{tab:moments} compares the first four moments of all
three distributions in Fig.~\ref{fig06}c.
While NN reproduces the mean more accurately, the standard deviation
is much better modelled with the stochastic approach. In other words,
while the average behaviour is better modelled with NN, for properly
reproduced the amplitude of tower vibrations the stochastic approach
is better suited.

Curiously however,
while the (weak) skewness is also better approximated by
the stochastic approach, the value of the kurtosis is better
reproduced by the NN model.

The performance of both models was also assessed using the metrics
introduced above, Eqs.~(\ref{eq:MAE})-(\ref{eq:SDofSE}) and results
are given in Tab.~\ref{tab:metrics}:
one clearly sees a tendency for lower
values of 
$\mathit{MAE}$, 
$\mathit{SDofAE}$, 
$\mathit{MSE}$ and 
$\mathit{SDofSE}$.
For such metrics, one concludes that the stochastic approach results have
comparable accuracy or are even more accurate than the results from
NN models, in what concerns the predictions for the tower top
acceleration.

As for the temporal correlations of the signals, the results plotted
in Fig.~\ref{fig07} show the spectral behaviour of the two models
compared with the measurements of the tower acceleration.
While NN better reproduces the periodic models (picks in the spectrum)
present in the measurements, it is worse for the low-frequency range
and for the overall amplitudes of the spectrum.
Notice the different vertical scale used for plotting the
NN results.
The stochastic approach smooths out the periodic modes, but reproduces
the overall shape of the empirical power spectrum.

Finally, a better reconstruction of two-point statistics is also obtained
through the stochastic approach, as shown in Fig.~\ref{fig08}.
It is clear that the stochastic approach better reconstructs the
differences $\Delta a$ between consecutive values in the tower
acceleration series, only with some small deviations for the largest
differences.
This shows furthermore that the assumption of Markovian process
with Langevin noise is considerably reliable for these data sets. 



\section{Conclusions}
\label{sec:conclusions}

We have presented a stochastic approach which is capable of modelling the normal behaviour of wind turbine tower acceleration for its monitoring. Our results demonstrate that, compared to neural network (NN) models and using a single input, the Langevin model is able to better reconstruct non-Gaussian fluctuations of signals. In general both models provide a good estimate for the central part of the original signal, but the stochastic approach also reconstructs its complete variance.

We have also shown that a normal behaviour model based on the stochastic approach provides more information about the original
signal than one based on NNs.
Still, one should emphasize that
previous publications on normal behaviour models, to monitor vibration of wind turbines based on SCADA \cite{Kusiak2010h, Zhang2012b} using 10-second sampling signals, demonstrated the suitability of NNs and control charts as a method to monitor signals in time-domain. Our results for NN match the applications found in the literature, where they have been used with 10-minute sampling time-series \cite{Schlechtingen2011}. When using low-frequency data, the NN is as good as our stochastic approach. 

Putting our results in perspective, we state that NNs are a good choice to predict average values or low fluctuating signals. For situations where $10$-minute data is not sufficiently sampled, namely when dealing with cumulative loads for which the large wind fluctuations of short intervals play an important role. In these situations, since the stochastic approach depends on the conditional limit of the first two moments in the data, when the time-lag between measurements tends to zero, it is better suited for signals with a higher sampling rate.

All in all, in those cases where the sampling ratio is low, the NNs remain as a good option to take into account, as the monitoring is sufficient in time-domain and the requirements to apply the stochastic approach are not always fulfilled. In cases where the sampling ratio of the signal is high ($1$ Hz in our case), the stochastic approach is clearly a better choice. It allows for monitoring normal behaviour in time-domain as with NNs, but also extends the possibilities to the frequency-domain.
To notice that a central advantage of the Langevin approach is that deterministic and noisy contributions can be separated leading to improvements in the statistics reproduction of the output fluctuations, since with the noisy contribution as a numerical function, quite general heavytailed distributions can be grasped.
The study of the long time behavior of the deterministic part of the Langevin equation used in the stochastic approach can also provide information about the health of the structure, in a similar direction as previous work \cite{rinn2013}. From this paper, one can now use the available routines \cite{philip,routine} and extend the stochastic approach here applied to other wind turbine properties, in particular to other loads. 

\section*{Acknowledgment}

This work was partly funded by the German Federal Ministry of Economic
Affairs and Energy and the State of Lower Saxony as part of the
research project "Probabilistic
load description, monitoring, and reduction for the next generation of
offshore wind turbines (OWEA Loads)", grant number 0325577B, and also
by the Ministry of Science and Culture of Lower Saxony in the project
``Ventus Efficiens'' (ZN3024).
The authors also thank Senvion SE for providing the data here analyzed.

\section*{References}

\bibliography{lvt_library.bib}

\end{document}